\documentclass[journal=ancac3,manuscript=article,layout=twocolumn]{achemso}

\usepackage[version=3]{mhchem} % Formula subscripts using \ce{}
\usepackage{bm}% bold math
\usepackage{amssymb,amsmath}
\usepackage{epstopdf}
\usepackage{graphicx} %graphics handler

\author{William D.~Rice}
\affiliation{Department of Electrical and Computer Engineering, Rice University, Houston, Texas 77005, USA}
\alsoaffiliation{Department of Physics and Astronomy, Rice University, Houston, Texas 77005, USA}

\author{Ralph T.~Weber}
\affiliation{Bruker BioSpin Corporation, Billerica, Massachusetts, 01821 USA}

\author{Ashley D.~Leonard}
\affiliation{Department of Chemistry, Rice University, Houston, Texas 77005, USA}

\author{James M.~Tour}
\affiliation{Department of Chemistry, Rice University, Houston, Texas 77005, USA}
\alsoaffiliation{Department of Mechanical Engineering and Materials Science, Rice University, Houston, Texas 77005, USA}

\author{Pavel Nikolaev}
\author{Sivaram Arepalli}
\affiliation{Department of Energy Science, Sungkyunkwan University, Suwon 440-746, South Korea}

\author{Vladimir Berka}
\author{Ah-Lim Tsai}
\affiliation{University of Texas Medical School, Houston, Texas 77005, USA}

\author{Junichiro Kono}
\email{kono@rice.edu}
\affiliation{Department of Electrical and Computer Engineering, Rice University, Houston, Texas 77005, USA}
\alsoaffiliation{Department of Physics and Astronomy, Rice University, Houston, Texas 77005, USA}

\title[\texttt{achemso} demonstration]
{Enhancement of the Electron Spin Resonance of Single-Walled Carbon Nanotubes\\
by Oxygen Removal}

\begin{document}
%%%%%%%%%%%%%%%%%%%%%%%%%%%%%%%%%%%%%%%%%%%%%%%%%%%%%%%%%%%%%%%%%%%%%
%% The manuscript does not need to include \maketitle, which is
%% executed automatically.  The document should begin with an
%% abstract, if appropriate.  If one is given and should not be, the
%% contents will be gobbled.
%%%%%%%%%%%%%%%%%%%%%%%%%%%%%%%%%%%%%%%%%%%%%%%%%%%%%%%%%%%%%%%%%%%%%
\begin{abstract}
We have observed a nearly fourfold increase in the electron spin resonance (ESR) signal from an ensemble of single-walled carbon nanotubes (SWCNTs) due to oxygen desorption.  By performing temperature-dependent ESR spectroscopy both before and after thermal annealing, we found that the ESR in SWCNTs can be reversibly altered via the molecular oxygen content in the samples.  Independent of the presence of adsorbed oxygen, a Curie-law (spin susceptibility $\propto 1/T$) is seen from $\sim$4~K to 300~K, indicating that the probed spins are finite-level species.  For both the pre-annealed and post-annealed sample conditions, the ESR linewidth decreased as the temperature was increased, a phenomenon we identify as motional narrowing.  From the temperature dependence of the linewidth, we extracted an estimate of the intertube hopping frequency; for both sample conditions, we found this hopping frequency to be $\sim$100~GHz.  Since the spin hopping frequency changes only slightly when oxygen is desorbed, we conclude that only the spin susceptibility, not spin transport, is affected by the presence of physisorbed molecular oxygen in SWCNT ensembles. Surprisingly, no linewidth change is observed when the amount of oxygen in the SWCNT sample is altered, contrary to other carbonaceous systems and certain 1D conducting polymers.  We hypothesize that physisorbed molecular oxygen acts as an acceptor ($p$-type), compensating the donor-like ($n$-type) defects that are responsible for the ESR signal in bulk SWCNTs.
\end{abstract}

\pagebreak

\section{Introduction}
%\noindent
One of the most important fields in modern physics is dedicated to understanding spin dynamics in condensed matter systems~\cite{Si1997PRL, Si1998PRL, BalentsEgger2000PRL, BalentsEgger2001PRB, KiselevPRB2000, RabelloSi2000EurophysLett, DeMartinoPRL2002, DoraPRL2008} and applied devices~\cite{DiVincenzo1995Science, Wolf2001Science}.  Spin transport is a sensitive probe of many-body correlations, as well as an indispensable process for spintronics.  When spins are dimensionally confined, especially to one dimension (1D), they are predicted to show strong correlations~\cite{BalentsEgger2000PRL, BalentsEgger2001PRB, RabelloSi2000EurophysLett, DeMartinoPRL2002, DoraPRL2008, Giamarchi2004Book} and long coherence times~\cite{KiselevPRB2000}.  Single-walled carbon nanotubes (SWCNTs) are ideal materials for studying 1D spin physics due to their long mean free paths and weak spin-orbit coupling~\cite{Ando2000JPSJ}.  Exotic spin properties in metallic SWCNTs at low temperatures and high magnetic fields have been predicted, including the appearance of a peak splitting in the spin energy density spectrum, which can be used to probe spin-charge separation in Luttinger-liquid theory~\cite{RabelloSi2000EurophysLett, DeMartinoPRL2002, DoraPRL2008}.

%%% FIG. 1 %%%
\begin{figure} [htbp]
\includegraphics [width=3.25 in] {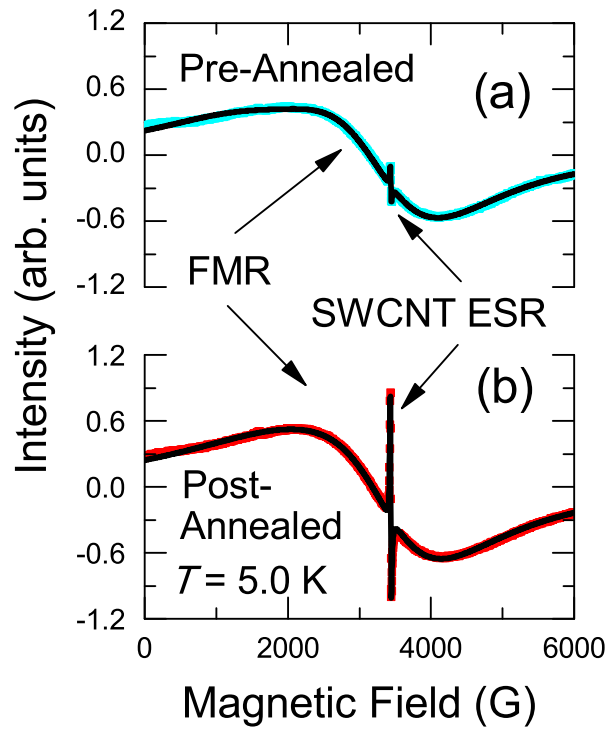}
\caption{\small \bf (a)  Full range ESR scan at 5.0~K of SWCNT sample before annealing (cyan), where the ESR signal is buried in the large FMR background.  (b)  Full scan of SWCNT sample at 5.0~K after annealing (red), where the SWCNT ESR is the dominant feature.  Black curves indicate fits composed of two large linewidth Lorentzian lines, which describe the FMR background, and a Dysonian line describing the SWCNT ESR.  The FMR intensities stay the same before and after annealing.}
\label{Long_scan_plot}
\end{figure}

Despite the predictions of unique and exciting spin-related phenomena, the spin properties of SWCNTs are still poorly understood.  One of the most ubiquitous experimental methods for studying spin dynamics, electron spin resonance (ESR) spectroscopy, has been the prime method used to measure the spin dynamics of SWCNTs, since it provides information on spin-orbit coupling, phase relaxation time, spin susceptibility, and spin diffusion.  However, there are substantial disagreements in the literature on SWCNT ESR linewidth, spin susceptibility, $g$-factor, and lineshape~\cite{KosakaCPL1994, KosakaCPL1995, PetitPRB1997, ClayePRB2000, SalvetatPRB2005, NafradiPSS2006, CorziliusPSSB2008, LikodimosPRB2007, AngladaPSSB2010, Rice2011}.  In particular, the temperature dependence of the spin susceptibility of SWCNTs has been surprisingly difficult to reproduce~\cite{PetitPRB1997, SalvetatPRB2005, LikodimosPRB2007, Rice2011}.  The experimental evidence is so conflicting that even the origin of the ESR in SWCNTs is under substantial uncertainty, with certain authors claiming it results from defects~\cite{KosakaCPL1995, SalvetatPRB2005, MussoDiaRelMater2006, AngladaPSSB2010} and others suggesting it is intrinsic to nanotubes~\cite{PetitPRB1997, ClayePRB2000, NafradiPSS2006, LikodimosPRB2007, KombarakkaranCPL2008, CorziliusPSSB2008}.

%%% FIG. 2 %%%
\begin{center}
\begin{figure*} [htbp]
\includegraphics [width = 6.5 in] {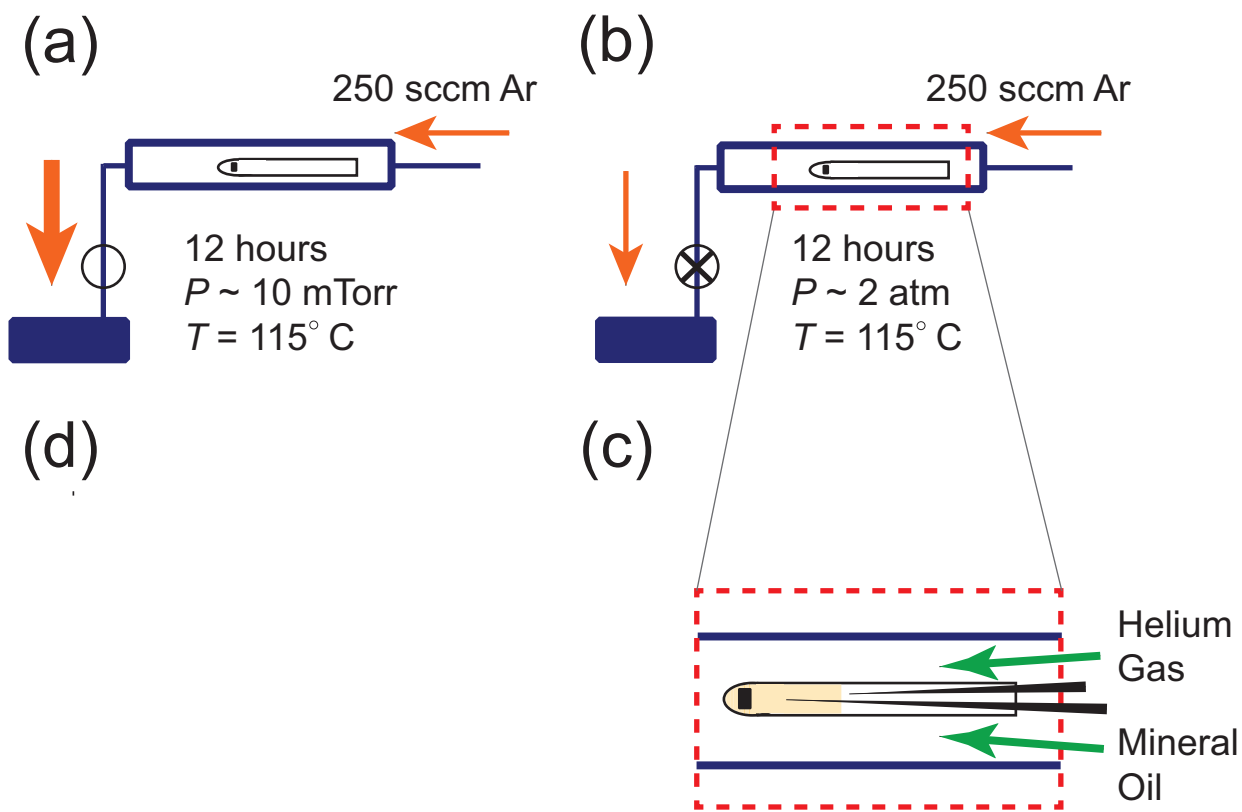}
\caption{\small \bf The steps for annealing the sample are given in (a)--(c).  (a)  First 12 hour anneal stage with $T$=115$^{\circ}$C and pressure held at 10~mTorr with a 250~sccm argon flow.  (b)  Second 12~hour anneal stage.  For this step, we raised the pressure inside the furnace to 2~atm with the vacuum valve held open just enough to keep the pressure steady.  (c)  To insure that oxygen did not re-contaminate the sample during ESR measurements, we backfilled the ESR tube with degassed mineral oil while streaming helium gas before the sample was removed from the furnace. (d)  A picture of the sample before it was placed inside the ESR tube.}
\label{Sample_prep_plot}
\end{figure*}
\end{center}

To complicate matters further, adsorbed molecular gas species, such as oxygen and hydrogen, have long been known to strongly affect spins in carbonaceous systems~\cite{PastorPR1956, ArnoldCarbon1967, SingerApplSpect1982, VittorioJMatRes1993, AtsarkinJMagRes2001, BardelebenDiaRelMat2003}.  ESR studies done on adsorbed gas species in carbon nanotubes have focused solely on adsorption of hydrogen~\cite{ShenPRB2003, ClewettPSSB2006, MussoDiaRelMater2006, KombarakkaranCPL2008}.  Interestingly, adsorbed hydrogen produces an \textit{increase} in the multi-walled carbon nanotube (MWCNT) ESR, while creating a \textit{decrease} in the ESR signal in SWCNTs~\cite{ShenPRB2003, ClewettPSSB2006}.  These diametrically opposing trends are not currently understood, especially given the fact that the origin of the SWCNT ESR is still under considerable debate.  However, many experimental studies have postulated that the signal from MWCNTs occurs from defects~\cite{ShenPRB2003, ClewettPSSB2006, MussoDiaRelMater2006, KombarakkaranCPL2008}, suggesting that hydrogen may not affect the intrinsic nanotube response.  %Several groups have utilized high temperature annealing of SWCNT samples to remove extraneous carbonaceous material~\cite{KosakaCPL1995, PetitPRB1997, NafradiPSS2006}, but a rigorous investigation of the role of adsorbed gas species has not been performed.

Here, we show that adsorbed molecular oxygen has a considerable influence on the spin susceptibility of SWCNT ensembles, while having only a very small impact on spin movement.  By looking at the ESR of SWCNTs as a function of temperature ($T$) both before (pre-annealed) and after (post-annealed) thermal annealing, we were able to quantitatively evaluate the impact of adsorbed oxygen spins in SWCNTs.  Strikingly, we found that oxygen desorption increased the ESR signal by nearly a factor of four.  When oxygen was added back to the system, the ESR signal was once again quenched, indicating reversibility.  Despite the change in signal intensity with oxygen, the spin hopping frequency stayed around $\sim$100~GHz for both annealing sample conditions. We hypothesize that the ESR SWCNT signal is due to $n$-type defects, which are compensated by $p$-type oxygen acceptor states when present in the sample.

The SWCNT sample consisted of acid-purified laser-oven nanotubes obtained from NASA~\cite{NikolaevJPCC2007}.  Despite the soft-bake acid-purification procedure used to remove ferromagnetic catalyst particles (cobalt and nickel), it is evident using ESR spectroscopy that they remain in the sample (\ref{Long_scan_plot}).  We performed both thermogravimetric analysis (TGA) and x-ray photoelectron spectroscopy (XPS) to determine the mass concentration of the metallic catalyst particles.  Using XPS, we determined the ratio of the Co to Ni catalyst to be 0.9:1.0, which is very close to the values reported by Nikolaev {\it et al}.~\cite{NikolaevJPCC2007}.  Subsequently, this ratio enables us to compute how much of the resulting oxidized mass produced by TGA (Figure S2 in the Supplementary Information) is due to the catalytic metal.  For this particular laser-oven sample, we attribute 5.8\% of the mass to ferromagnetic catalyst particles.  This non-negligible mass percentage explains why the ferromagnetic resonance (FMR) from the catalyst particles dominates the background of the ESR scans of SWCNTs, as seen in~\ref{Long_scan_plot}~(a).

%%% FIG. 3 %%%
\begin{center}
\begin{figure*} [ht]
\includegraphics [width=6 in] {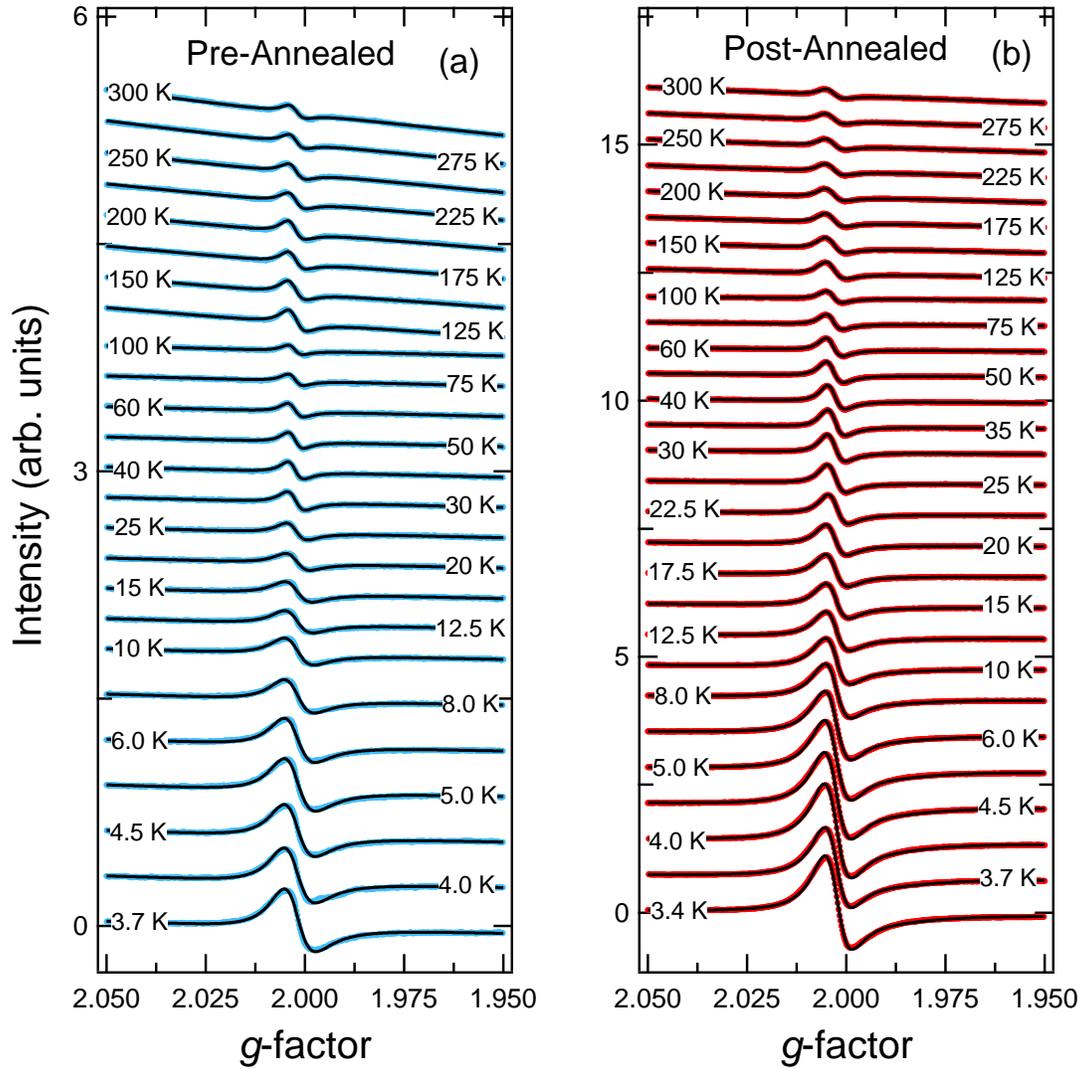}
\caption{\small \bf Raw ESR scans of sample (a) before (cyan curves) and (b) after (red curves) annealing as a function of $T$. Black lines indicate fits to the traces.}
\label{Short_scan_plot}
\end{figure*}
\end{center}

The SWCNT sample used for ESR measurements was prepared with meticulous attention to compaction and sample homogeneity, since small perturbations in the SWCNTs position affect the ESR signal (details presented in the Materials and Methods section).  A pictorial scheme, given in~\ref{Sample_prep_plot}, shows the steps taken to anneal the sample after it has been created.  The powdered SWCNT ensemble was compacted into a pellet and then placed in a dessicator for several days to create a water-free SWCNT pellet; this sample condition is denoted in the text as ``pre-annealed.''  The 0.6~mg (0.38~g/cm$^3$) SWCNT pellet was then put into a quartz ESR tube and held in place with quartz wool.  X-band (9.6~GHz) temperature-dependent ESR measurements were taken at discrete temperatures from 3.7~K to 300~K.  The sample was then brought to a vacuum furnace and annealed at 115$^{\circ}$C for 24~hours.  After the annealing, the ESR tube was then rigorously prepared to prevent re-exposure to air.  In the text, we refer to this sample condition as ``post-annealed.''  As with the pre-annealed sample condition, temperature dependent ESR scans were performed on the post-annealed sample from 3.4~K to 300~K.

\section{Results}
\noindent
~\ref{Long_scan_plot} shows the remarkable difference that occurs upon thermal annealing:  the SWCNT ESR peak goes from being buried in the large catalyst FMR background to dominating the spectrum.  Despite the tremendous increase in the SWCNT ESR signal, the FMR hardly changes, demonstrating that the annealing \textit{only} affects the nanotube ESR.  The full scan spectra were fit using three curves:  two large-linewidth Lorentzians for the two catalyst species, Co and Ni, and a Dysonian line for the SWCNT ESR centered at 3430~G when the microwave frequency is 9.6~GHz~\cite{Rice2011}.  No arbitrary slope or numerical offsets were used in addition to the three lines.  The long scan fits, represented by the black curves in~\ref{Long_scan_plot}, describe the observed spin resonance spectrum very well.

A more detailed set of scans covering two orders of magnitude in temperature, which are shown in~\ref{Short_scan_plot}, allow a more quantitative understanding of the SWCNT ESR.  The raw data traces presented in~\ref{Short_scan_plot} are fit with a Dysonian lineshape (black traces) for each different temperature.  The baseline, which becomes more prominent at higher temperatures, is simultaneously fit using two large linewidth Lorentzians to account for the slowly varying FMR background.  The lack of secondary peaks and anomalous shapes, as is seen in other publications~\cite{KosakaCPL1994, SalvetatPRB2005, AngladaPSSB2010}, attests to the purity of our sample.  If MWCNTs~\cite{BeuneuPRB1999}, amorphous carbon~\cite{DemichelisDRM1994}, or graphite~\cite{WagonerPR1960} contributed to the ESR signal, they would each manifest themselves differently in lineshape and as a function of temperature.  Therefore, we conclude that we are observing a spin resonance from SWCNTs.  ~\ref{Short_scan_plot} shows that for both sample conditions, the SWCNT ESR increases as $T$ decreases.  Since the signal intensity is proportional to the mass spin susceptibility, $\chi_{\rm g}$, the ESR signal change as a function of $T$ strongly suggests that we are not observing Pauli law behavior, in which $\chi_{\rm g}$ is independent of $T$.  Further, as seen in both~\ref{Long_scan_plot} and~\ref{Short_scan_plot}, the ESR signal size changes dramatically upon annealing.

%%% FIG. 4 %%%
\begin{figure} [htbp]
\includegraphics [width=3.5 in] {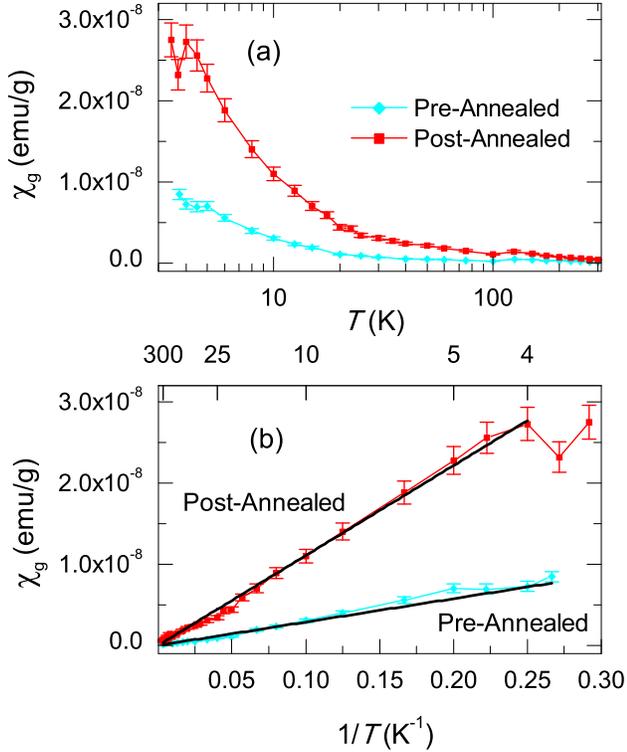}
\caption{\small \bf (a) Mass spin susceptibility $\chi_{\rm g}$ as a function of $T$ for the sample before annealing (cyan) and after annealing (red). (b)~$\chi_{\rm g}$ versus $1/T$ for the sample before annealing (cyan) and after annealing (red). Annealing produces a large increase in the magnitude of $\chi_{\rm g}$, but does not change its temperature dependence.  Both sets of data can be fit well using a Curie-law model (black lines) down to $\sim$4~K.}
\label{Chi_plot}
\end{figure}

To examine these observations more quantitatively, we extracted $\chi_{\rm g}$ from the fitted spectra in~\ref{Short_scan_plot}.  We used a CuSO$_4\cdot$5H$_2$O spin concentration standard measured at temperatures between 4~K and 300~K to obtain mass spin susceptibility values from our measurements.  \ref{Chi_plot} shows how $\chi_{\rm g}$ varies as a function of $T$ for both sample conditions.  The large increase in $\chi_{\rm g}$ between the pre-annealed and post-annealed sample conditions is striking, especially at low temperatures.  Clearly, the annealing process greatly augments the number of spins probed.

Plotting $\chi_{\rm g}$ versus $1/T$ shows that for both the pre and post-annealed sample conditions, the spin susceptibility follows a Curie-law behavior:
\begin{equation}
\chi_{\rm g} = C/T,
\label{Curie-law equation}
\end{equation}
\noindent where $C$ is the Curie constant and is equivalent to $\frac{\mu_B^2 N}{k_B}$, $\mu_B$ is the Bohr magneton, $N$ is the number of probed spins, and $k_B$ is the Boltzmann constant.  For the pre-annealed sample condition, we fit the values of $\chi_{\rm g}$ from 300~K to 3.7~K and find that $C$ = 2.88$\pm$0.21$\times10^{-8}$~emu-K/g, which translates into $N = 7.9\times10^{12}$~spins.  This value for the pre-annealed Curie constant is very close to what Likodimos \textit{et al.} found in their work~\cite{LikodimosPRB2007}, as well as what other groups have estimated their mass susceptibility to be~\cite{SalvetatPRB2005, NafradiPSS2006}.  Similarly, if we fit the post-annealed data from 300~K to 4.0~K, we extract $C$ = 1.11$\pm$0.04$\times10^{-7}$~emu-K/g (3.0$\times10^{13}$~spins).  Therefore, by removing adsorbed gases from the SWCNTs, the number of spins probed is increased by a factor of 3.9.  Interestingly, both pre-annealed and post-annealed sample conditions show Curie-law trends down to $\sim$4~K.  However, as the temperature drops below 4~K, $\chi_{\rm g}$ seems to deviate from~\ref{Curie-law equation}, especially for the post-annealed sample; further investigation is needed to determine how $\chi_{\rm g}$ varies with $T$ at temperatures below 4~K.

%%% FIG. 5 %%%
\begin{figure} [htbp]
\includegraphics [width=3.25 in] {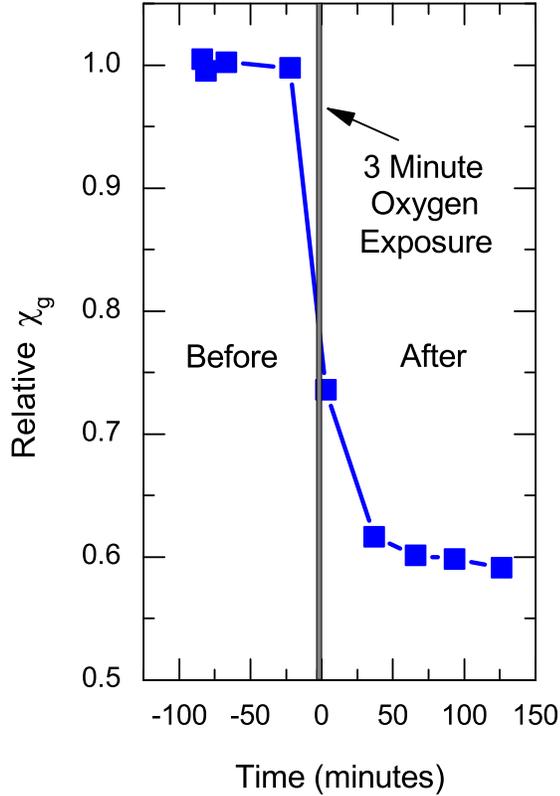}
\caption{\small \bf Relative spin susceptibility of an annealed SWCNT sample over several hours shows a sudden decrease once oxygen is introduced (gray shaded region).  After nearly two hours, the decrease in spin susceptibility appears to stop, indicating that equilibrium has been reached.}
\label{Oxygen_plot}
\end{figure}

To investigate which adsorbed gas species might contribute to this marked rise in the spin susceptibility, we applied a constant pressure of pure oxygen onto a freshly annealed powdered SWCNT sample.  Immediately after the oxygen exposure, we ran an ESR scan to interrogate how the sample responded to the presence of oxygen.  We repeated the ESR scans every 30~minutes to see how the signal evolved (details given in the Supplementary Information).  As~\ref{Oxygen_plot} shows, the relative spin susceptibility dramatically decreases as the oxygen is adsorbed onto the SWCNTs, dropping to 59\% of the initial annealed spin susceptibility.  As a comparison, the ratio of the pre-annealed spin susceptibility to the post-annealed spin susceptibility at $T = $300~K is 0.42.  The fact that the presence of oxygen does not fully return the ESR to its original suppressed value may be due to the inability of the oxygen to return to its former locations between SWCNTs. Once the oxygen is removed, the strong van der Waals forces between tubes reduce the intertube spacing, thus preventing oxygen from quenching the ESR.  This test unequivocally shows that oxygen is a strong spin suppressor in SWCNTs.  Importantly,~\ref{Oxygen_plot} demonstrates that the annealing effect is \textit{reversible}:  we can remove molecular oxygen and increase $\chi_{\rm g}$, and then add molecular oxygen back to the system and suppress $\chi_{\rm g}$.

%%% FIG. 6 %%%
\begin{center}
\begin{figure*} [ht]
\includegraphics [width=6.5 in] {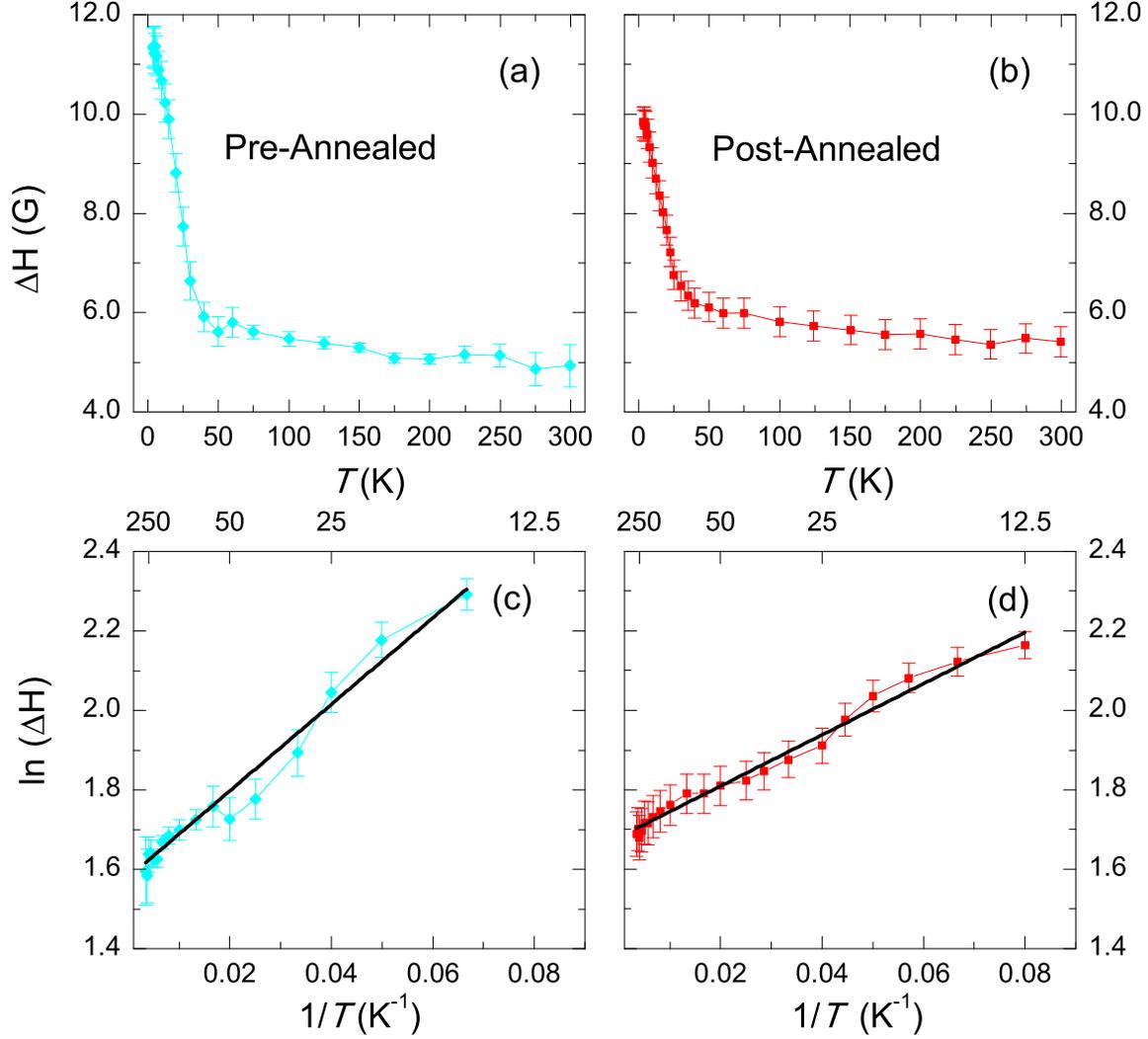}
\caption{\small \bf ESR linewidth versus $T$ for the (a) pre-annealed and (b) post-annealed sample conditions. In (c) and (d), the natural logarithm of the data in (a) and (b) are plotted versus $1/T$ down to 15~K and 12.5~K, respectively. Black lines indicate linear fits to ln$(\Delta H)$ versus $1/T$ data from which we can extract intertube hopping frequencies.}
\label{Linewidth_plot}
\end{figure*}
\end{center}

\section{Discussion}
\noindent
~\ref{Short_scan_plot} shows that, in addition to the ESR signal size increasing as $T$ decreases, the half-width at half-maximum of the line, $\Delta H$, becomes larger as $T$ is lowered.  This phenomenon, which we have previously reported on~\cite{Rice2011}, is known as motional narrowing~\cite{BloembergenPR1948,KuboJPhysSocJapan1954}.  Interestingly, we observe that this behavior occurs for both the annealed and non-annealed sample conditions.

We can see the phenomenon of motional narrowing more clearly in~\ref{Linewidth_plot}.  The exponential rise in $\Delta H$ is well described by a simple activation model
%
%%%%%  EQ. 1  %%%%%
\begin{equation}
\Delta H = \Delta H_0\times \textrm{exp}\left(\frac{\Delta E}{k_{\rm B}T}\right),
\label{linewidth equation}
\end{equation}
%%%%%%%%%%%%%%%%%%%%%
%
where $\Delta H_0$ is the high-$T$ (``metallic'') limit of the linewidth, $\Delta E$ is the energy required to move from one location to another, and $k_{\rm B}$ is the Boltzmann constant.  For the pre-annealed sample condition, the data from 15~K to 300~K was fit by~\ref{linewidth equation}, as shown in~\ref{Linewidth_plot}.  We extracted a $\Delta H_0$ of 4.85$\pm$0.04~G with a hopping energy of 0.938$\pm$0.036~meV ($T_{\rm hop}$ = 10.88~K or $\nu_{\rm hop}$ = 227~GHz) from the pre-annealed linewidth.  Similarly, we fit the post-anneal data from 12.5~K to 300~K, which yielded a hopping energy of 0.560$\pm$0.016~meV ($T_{\rm hop}$ = 6.50~K or $\nu_{\rm hop}$ = 135~GHz) with a $\Delta H_0$ of 5.35$\pm$0.03~G.

Quantitatively, the annealing procedure seems to have only a minor effect on the hopping energy, $\Delta E$, changing $T_{\rm hop}$ from 10.88~K to 6.44~K.  This small change in the hopping characteristics suggests that adsorbed gases have little effect on the motion of the spins. In a similar manner, annealing has a small effect on the linewidth in the high-$T$ limit, $\Delta H_0$.  Annealing increases $\Delta H_0$ by $\sim$0.5~G, which correspondingly changes the spin dephasing rate, $T_2$ $(T_2 = \frac{\hbar}{g\mu_{\rm B}}\frac{1}{\Delta H})$.  Before annealing, the high-$T$ value of $T_2$ is 11.7~ns, which drops slightly to 10.6~ns after annealing.  As with the hopping energy changes, the fact that annealing has only a minor effect on $\Delta H_0$ indicates that adsorbed gases in bulk SWCNT systems have little impact on the mobility of the probed spins.

The fitting to the linewidth should be viewed as a quantitative estimate, since the entire temperature dependence is not described by~\ref{linewidth equation}.  Nevertheless, the fundamental narrowing behavior of $\Delta H$ is well captured for most of the investigated temperature range.  If clear differences between the pre-annealed and post-annealed linewidths existed, they should be understood by our simple activation model.  Other, more complicated theoretical frameworks, such as those used to describe similar physical behavior seen in 1D conducting polymers, may also be good fits to the data (see, e.g., work on 1D conducting polymers~\cite{NechtscheinPRB1983,MizoguchiPRB1995}).  Unfortunately, clear and precise knowledge of the spin diffusion rate (both perpendicular and parallel to the tube axis) is needed, which has not been measured as a function of $T$ for SWCNTs.  

%%% FIG. 7 %%%
\begin{figure} [htbp]
\includegraphics [width=3.25 in] {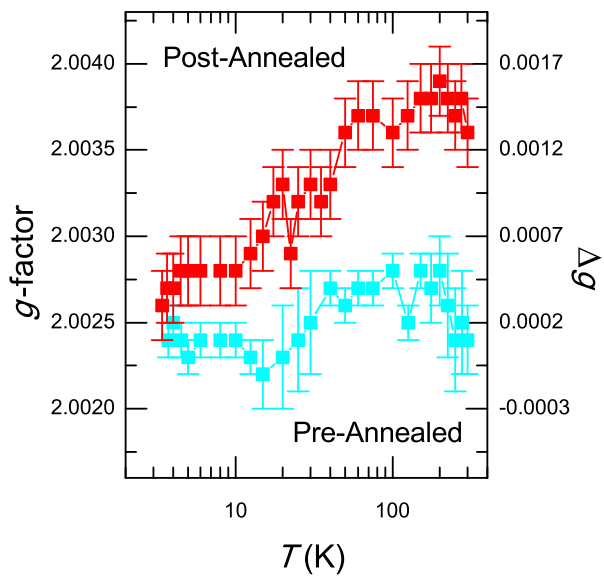}
\caption{\small \bf Experimentally obtained $g$-factor values as a function of $T$ for the sample before (cyan) and after (red) annealing.}
\label{g-factor_plot}
\end{figure}

As mentioned above, the mobility of the spins is marginally influenced by the presence of adsorbed gases; rather, it seems to be strongly dependent on the sample temperature, as seen in~\ref{Linewidth_plot}.  The center position of the resonance, $H_0$, on the other hand, is only slightly dependent on the temperature (\ref{g-factor_plot}).  As is typical in magnetic resonance experiments, we define the $g$-factor as the ratio of the perturbing field energy to the magnetic dipole energy $g = \frac{h\nu_0}{\mu_B H_0}$, where $h$ is Planck's constant and $\nu_0$ is the frequency of the perturbing microwave field.  The measured $g$-factor for the pre-annealed sample condition is nearly temperature independent, staying close to the free electron $g$-factor value of 2.0023 throughout the full 3.7~K to 300~K temperature range.  This behavior is similar to the temperature dependence of the $g$-factor of graphite in the plane perpendicular to the $c$-axis~\cite{WagonerPR1960, MatsubaraPRB1991}.  The $g$-factor of the post-annealed sample condition shows a slight increase with increasing temperature.  However, there is a small, but measurable, change in the $g$-factor upon annealing, especially at high temperatures.  If we take $g_0$ to be the free electron $g$-factor (2.0023), we can define the quantity $\Delta g \equiv g-g_0$.  Following Platzman~\cite{PlatzmanSSP1973}, we interpret $\Delta g$ to be proportional to the spin-orbit coupling.  As such, the upward shift ($\Delta g$ changes by $\sim$0.001 at $T$ = 300~K) in the $g$-factor upon annealing indicates that adsorbed gas species weakly quench some of the spin-orbit coupling of the probed spins.  %Last, from the data shown in~\ref{g-factor_plot}, we can rule out the possibility that the Elliott-Yafet mechanism is responsible for the spin relaxation in our system~\cite{YafetPR1952, ElliottPR1954}, since we would expect as $\Delta H$ becomes larger (shorter value of $T_2$), $\Delta g$ would become larger.

As~\ref{Oxygen_plot} shows, the quenching of the spin susceptibility is reversible when molecular oxygen is introduced.  Given this information, as well as the fact we are using a low annealing temperature of 115$^{\circ}$C (33~meV) to create the changes in $\chi_{\rm g}$, it seems most likely that the spin susceptibility change is caused by physisorbed molecular oxygen:  that is, no change in the chemical bonds of the system occurs when oxygen is either added or removed.  Indeed, this line of reasoning follows Ulbricht {\it et al}., who showed that molecular oxygen primarily physisorbs to SWCNTs~\cite{UlbrichtPRB2002} (unlike reactive atomic oxygen, which is known to form epoxies and ethers in SWCNTs~\cite{LarcipreteCarbon2009}).  In addition, theoretical work has also shown that molecular oxygen should physisorb to SWCNTs~\cite{GiannozziJChemPhys2003, DagCPL2003, DagPRB2003}.

Taken as a whole, our results show that molecular oxygen physisorbed onto SWCNTs affects only $\chi_{\rm g}$, with minimal changes to $\Delta H$ and the $g$-factor.  The lack of change in $\Delta H$ is especially odd, since in many biological systems, O$_2$ serves as general relaxer for various paramagnetic species via spin-spin interaction, which leads to ESR line broadening. This very nature finds wide application of oxygen in membrane topology studies of membrane bound macromolecules~\cite{MillhauserMethEnz1995, MalmbergBioChem2003} and serves as the theoretical basis for oximetry in quantifying O$_2$ in a particular biological sample present in aqueous environment~\cite{AhmadChemRev2010}.  Furthermore, the observation that oxygen does not affect the observed $\Delta H$ in SWCNTs also contrasts with previous ESR stuides on gas adsorption in 1D polymer systems, such as \textit{trans}-polyacetylene, which show that $\Delta H$ increases as the number of gas molecules (pressure) increases.  For example, Houz\'{e} and Nechtschein~\cite{HouzePRB1996} show that exchange and dipolar interactions resulting from an adsorbed gas species broaden the ESR linewidth, $\Delta\omega$, with the relation given as $\delta(\Delta \omega) = p\omega_{\rm hop}C_i$.  Here $p$ is the probability of a spin flip due to an interaction, $\omega_{\rm hop}$ is the hopping frequency of the probed spin species, and $C_i$ is the concentration of the adsorbed gas.  If this type of process was occurring in our system, we should see some change in $\Delta H$ $\left(\Delta H = \frac{\hbar\Delta\omega}{g\mu_B}\right)$ when the oxygen is desorbed (annealing) or adsorbed (oxygen exposure test); however, no significant change in $\Delta H$ is observed.  

A very fast relaxation process could exist in our system.  Presumably, this latent ESR signal would show up as a change in the baseline, assuming that the resonance width was on the order of 100 to 1000~G.  However, this ``spin segregation'' into two different relaxation processes seems unlikely, especially since the baseline changes only slightly when the system is exposed to oxygen (see Figure S2 in the Supplementary Information).

%**We can also rule out the possibility that we are observing a band structure modification from physisorbed oxygen~\cite{DagCPL2003, DagPRB2003}.  Under this scenario, we must assume that the SWCNT ESR comes from metallic nanotubes, since changes in the band structure would otherwise not effect the resonance.  Although we are observing a Dysonian lineshape indicating that the probed spins are moving, we clearly see a Curie-law behavior in the spin susceptibility, which strongly suggests that we are not probing itinerant electron spins.  In addition, we would na\"{\i}vely expect an Elliott-Yafet relaxation mechanism if the signal stemmed from metallic-like SWCNTs, which we also do not observe.

A more plausible hypothesis is that oxygen is passivating the paramagnetic moments of SWCNT defects.  These defects could be caused during SWCNT growth, processing, or even acid purification~\cite{NikolaevJPCC2007}.  As shown in the discrete Raman spectroscopy plot in the Supplementary Information (Figure S1), we have a measurable amount of lattice defects, so an ESR signal from, say topological defects or dangling bonds, is possible.  Despite the localized nature of the ESR active species, the unpaired electrons still have mobility in this scenario:  the Dysonian signal in our work does not come from conduction electrons, as with ESR from bulk metals, but from thermally-activated finite-level species.  In addition, if we were probing defects or dangling bonds, we would expect both a Curie-law spin susceptibility temperature dependence, as well as a $g$-factor that is close to the free electron value; we observe both behaviors in our system.  Additionally, ESR arising from non-pristine SWCNTs would explain why the tremendous variation in ESR signals exists in the literature:  in essence, different SWCNT purification methods and sample preparation steps would create differing resonance signals.  

We can quantify how prevalent these ESR-active defects are by using the extracted estimate for the Curie-coefficient of the post-annealed sample (1.11$\pm$0.04$\times10^{-7}$~emu-K/g).  If we assume that all of the SWCNTs in our sample are (10,10) nanotubes, we find that there is roughly one probed spin for every 3$\times10^5$ carbon atoms.  Though this number in itself may not establish that the ESR signal arises from non-pristine SWCNTs, it does suggest an exceedingly weak response (assuming a SWCNT length of 1~$\mu$m, $C$ = 3.60$\times10^{-25}$~emu-K/SWCNT).

The quenching of the signal from oxygen exposure may result from a compensation mechanism.  If, for example, the defects are responsible for the ESR signal are $n$-type, creating states closer to the conduction band edge within the band gap, then they can be compensated by the introduction of acceptors.  Oxygen is thought to be a $p$-type acceptor to SWCNT, especially in the presence of defects~\cite{GrujicicaApplSurfSci2003, SavageJPhysCondMat2003}.  If the oxygen molecule is in the spin-singlet state, even charge transfer (chemisorption) can occur~\cite{JhiPRL2000, GrujicicaApplSurfSci2003, SavageJPhysCondMat2003}.  The adsorption of oxygen molecules to the ESR active defects would suppress the ESR signal; subtracting weakly-bound oxygen from the system would reverse this quench.  A parametric study of ESR as a function of oxygen pressure is needed to further clarify our observations, including the exact nature of the defects.

In conclusion, we have clearly demonstrated that spin susceptibility in nanotubes is strongly influenced by the presence of physisorbed oxygen.  We observe a nearly fourfold increase in spin susceptibility by desorbing gases present in our SWCNT ensemble.  The presence of adsorbed gases is shown not to substantially affect the spin hopping rate of $\sim$100~GHz.  The $g$-factor is found to be very close to the free electron value for both the pre-annealed and post-annealed sample conditions.  The decrease of the $g$-factor with decreasing temperature is not well-explained.  We hypothesize that the spin suppression is due to compensation of donor-type defects by acceptor oxygen states.

\section{Materials and Methods}
The sample consisted of acid-purified laser-oven SWCNTs obtained from NASA~\cite{NikolaevJPCC2007}.  We used extreme care to prepare the SWCNT sample for ESR measurements, since both compaction density and particle looseness have an effect on the ESR signal.  The powdered SWCNT ensemble was gently bath sonicated (Cole-Parmer, Model B3-R) at 12~W and a frequency of 55~kHz.  Sonication lasted for two hours in water before the mixture was ultracentrifuged at 26000~rpm (an average force of 88000~g's using a Sorvall AH-629 rotor with 36~mL centrifuge tubes) for four hours.  After ultracentrifugation, the supernatant was removed, and the resulting SWCNT pellet was extracted from the centrifuge tube and placed into an open-air oven at 115$^{\circ}$C for 15 minutes to remove most of the water.  At this point, the pellet was dry enough to be handled as a solid unit.  It was then placed into a dessicator for several days to create a water free SWCNT pellet; this sample condition is denoted in the text as ``pre-annealed.''  The 0.6~mg (0.38~g/cm$^3$) SWCNT pellet was held in a 3~mm diameter quartz ESR tube using a quartz wool.

The sample was then placed into a vacuum furnace kept at 115$^{\circ}$C for 24~hours.  For the first 12~hours, the pressure inside the furnace was maintained at 10~mTorr vacuum with a 250~sccm flow of purified argon; for the last 12~hours, the argon pressure was increased to $\approx$2~atm pressure with a 250~sccm argon flow.  After the annealing, the ESR tube was partially filled with degassed mineral oil.  While the mineral oil was being inserted into the ESR tube, a helium gas flow was simultaneously being applied so as to create a positively pressurized helium blanket.  Both the mineral oil and the gas were introduced into the ESR tube while it lay in the furnace using 12" syringes.  Helium gas was blanketed on top of the mineral oil before the ESR tube was mechanically sealed to prevent exposure to air.  In the text, we refer to this sample condition as ``post-annealed.''

Temperature-dependent ESR measurements of the pellet were taken at discrete temperatures from $\sim$3~K to 300~K using a Bruker EMX X-band (9.6~GHz) spectrometer.  An Oxford ESP900 cryostat with a ITC503 temperature controller and GFS600 transfer line was used for temperatures from 3.4~K to 100~K, and a BVT3000 temperature controller with a silver-coated double-jacketed glass transfer line was used for temperatures above 100~K.  From 3~K to 100~K, 200~$\mu$W of microwave power was applied to the cavity; above 100~K, a microwave power of 1~mW was used.  To ensure that we were in the linear power regime, we performed power dependence from 6~$\mu$W to 200~mW at temperatures from 4~K to 100~K and from 50~$\mu$W to 200~mW at temperatures from 125~K to 300~K.  The ESR signal was not saturated until the power exceeded 10~mW, ensuring the measurements were performed within the linear regime.  In addition, we extract estimates of our measurement error from fitting results performed on spectra in the linear regime.

A 2,2-diphenyl-1-picrylhydrazyl (DPPH) standard was used to calibrate the field.  We used a 1~mM CuSO$_4\cdot$5H$_2$O solution as a spin concentration standard to extract a numerical estimate of $\chi_g$; its ESR signal was measured at numerous temperatures from 4~K to 300~K.  We placed the standard in the same configuration as the SWCNT sample with approximately the same microwave cavity volume to minimize differences between the sample and reference.

%% classes.  Rather than use \section, an appropriate macro is
%% provided that will always work.
%%%%%%%%%%%%%%%%%%%%%%%%%%%%%%%%%%%%%%%%%%%%%%%%%%%%%%%%%%%%%%%%%%%%%
\acknowledgement

This work was supported by the DOE/BES through Grant No.~DEFG02-06ER46308, the Robert A.~Welch Foundation through Grant No.~C-1509, the Air Force Research Laboratories under contract number FA8650-05-D-5807, the W.~M.~Keck Program in Quantum Materials at Rice University, and Korean Ministry of Education, Science and Technology under the World Class University Program (R31-2008-10029).  We thank Qimiao Si, Adilet Imambekov, and Robert Hauge for useful discussions.

%% \suppinfo macro.
%%%%%%%%%%%%%%%%%%%%%%%%%%%%%%%%%%%%%%%%%%%%%%%%%%%%%%%%%%%%%%%%%%%%%
%\begin{suppinfo}
%Detailed sample characterization data and further information on the analysis and experimental methods are available in the Supporting Information.
%\end{suppinfo}

%\bibliography{ESR-ii}

\providecommand*\mcitethebibliography{\thebibliography}
\csname @ifundefined\endcsname{endmcitethebibliography}
  {\let\endmcitethebibliography\endthebibliography}{}

\end{document}